\documentstyle[12pt,times,subeqnar]{article}
\input epsf.tex

\newcommand{\nc}{\newcommand}
\newcommand{\rnc}{\renewcommand}

\rnc{\baselinestretch}{1.24}	
\rnc{\arraystretch}{1}   	

\makeatletter
\rnc{\theequation}{\thesection.\arabic{equation}}
\@addtoreset{equation}{section}
\makeatother                      

\nc{\be}{\begin{equation}}
\nc{\ee}{\end{equation}}
\nc{\bea}{\begin{subeqnarray}}
\nc{\eea}{\end{subeqnarray}}
\nc{\xx}{\nonumber\\}

\nc{\eq}[1]{(\ref{#1})}
\nc{\newcaption}[1]{\centerline{\parbox{6in}{\caption{#1}}}}

\nc{\fig}[3]{
\begin{figure}
\centerline{\epsfxsize=#1\epsfbox{#2.eps}}
\newcaption{#3. \label{#2}}
\end{figure}
}

\nc{\mytable}[3]{
\begin{table}
\centerline{#2}
\newcaption{#3. \label{#1}}
\end{table}
}

\nc{\np}[3]{Nucl. Phys. {\bf B#1} (#2) #3}
\nc{\pl}[3]{Phys. Lett. {\bf #1B} (#2) #3}
\nc{\pr}[3]{Phys. Rev. {\bf #1} (#2) #3}
\nc{\prl}[3]{Phys. Rev. Lett.{\bf #1} (#2) #3}
\nc{\prd}[3]{Phys. Rev. {\bf D#1} (#2) #3}
\nc{\ap}[3]{Ann. Phys. {\bf #1} (#2) #3}
\nc{\prep}[3]{Phys. Rep. {\bf #1} (#2) #3}
\nc{\rmp}[3]{Rev. Mod. Phys. {\bf #1} (#2) #3}
\nc{\cmp}[3]{Comm. Math. Phys. {\bf #1} (#2) #3}
\nc{\mpl}[3]{Mod. Phys. Lett. {\bf #1} (#2) #3}
\nc{\cqg}[3]{Class. Quant. Grav. {\bf #1} (#2) #3}
\nc{\jhep}[3]{J. High Energy Phys. {\bf #1} (#2) #3}
\nc{\atmp}[3]{Adv. Theor. Math. Phys. {\bf #1} (#2) #3}

\def\CN{{\cal N}}

\def\IR{{\hbox{{\rm I}\kern-.2em\hbox{\rm R}}}}
\def\IB{{\hbox{{\rm I}\kern-.2em\hbox{\rm B}}}}
\def\IN{{\hbox{{\rm I}\kern-.2em\hbox{\rm N}}}}
\def\IC{\,\,{\hbox{{\rm I}\kern-.59em\hbox{\bf C}}}}
\def\IZ{{\hbox{{\rm Z}\kern-.4em\hbox{\rm Z}}}}
\def\IP{{\hbox{{\rm I}\kern-.2em\hbox{\rm P}}}}
\def\IH{{\hbox{{\rm I}\kern-.4em\hbox{\rm H}}}}
\def\ID{{\hbox{{\rm I}\kern-.2em\hbox{\rm D}}}}

\def\a{\alpha}
\def\b{\beta}
\def\ga{\gamma}
\def\d{\delta}
\def\ep{\epsilon}

\def\l{\lambda}
\def\m{\mu}
\def\n{\nu}

\def\s{\sigma}
\def\t{\tau}

\def\G{\Gamma}

\def\half{\textstyle{\frac{1}{2}}} 
\def\imp{\Longrightarrow}

\def\del{\nabla}

\def\dirac{\nabla\!\!\!\!{/}}
\def\identity{1}

\nc{\group}[1]{\bigskip\noindent\underline{Group {#1}}\bigskip}
\nc{\recip}[1]{\textstyle{1\over #1}}

\begin{document}

\begin{titlepage}

\begin{flushright}
hep-th/9906105\\
KIAS-P99038\\
\end{flushright}
\vspace*{2.0cm}
\centerline{\Large\bf Mass Spectrum of $\mathbf{D=11}$ Supergravity} 
\vspace*{0.5cm}
\centerline{\Large\bf on AdS$\mathbf{_2\times S^2\times T^7}$}
\vspace*{1.5cm} 
\centerline{Julian Lee and Sangmin Lee
\footnote{jul, sangmin@kias.re.kr}}
\vspace*{1.0cm}
\centerline{\sl School of Physics}
\centerline{\sl Korea Institute for Advanced Study}
\centerline{\sl Seoul, 130-012, Korea}
\vskip0.3cm
\vspace*{2.0cm}
\centerline{\bf ABSTRACT}
\vspace*{0.5cm}
\noindent
We compute the Kaluza-Klein mass spectrum of the $D=11$ supergravity
compactified on $AdS_2\times S^2\times T^7$ and arrange it into 
representations of the $SU(1,1|2)$ superconformal algebra.
This geometry arises in M theory as the near horizon limit of
a $D=4$ extremal black-hole 
constructed by wrapping four groups of M-branes along the $T^7$. 
Via AdS/CFT correspondence, our result gives a prediction for
the spectrum of the chiral primary operators in the 
dual conformal quantum mechanics yet to be formulated.

\vspace*{1.1cm}

\end{titlepage}
\setcounter{footnote}{0}


\section{Introduction}

Among all known examples of the $AdS$/CFT correspondence 
\cite{malda, gkp, witt, rev}, 
the least understood is the $AdS_2$/CFT$_1$ case. 
The $D=1$ conformal field theory (CFT), or conformal quantum mechanics (CQM), 
has not been formulated and therefore no quantitative comparison
between the two sides of the duality has been made. 
See \cite{gibtown, town} for proposals on the CQM and 
\cite{strom, frag, nak, cad} for progress made in the bulk theory.

One of the most elementary check of the correspondence is to 
compare the spectrum of the two theories. In particular, the Kaluza-Klein (KK) 
mass spectrum of the supergravity (SUGRA) on $AdS$ is identified with the 
spectrum of chiral primary operators in the dual CFT.
One may hope that the KK spectrum of a SUGRA on $AdS_2$ may give a clue to 
formulate the dual CQM. 

The goal of this paper is to compute the KK spectrum in the cases where 
the $AdS_2$ is part of a string/M theory vacuum.
We specialize in the example of $D=11$ SUGRA 
compactified on $AdS_2\times S^2\times T^7$. 
\footnote{We thank Seungjoon Hyun for bringing our attention to this 
example.} We consider only the zero modes in $T^7$. 
{}From the string theory point of view, this theory is a valid approximation 
when  $R >> r, \tilde r  $, where $r,\tilde r$ are the radius 
and the dual radius of $T^7$ respectively, 
and $R$ is the radius of the sphere. 
In what follows, we will put $R$ to $1$ for simplicity. 
To obtain this geometry from M theory, one begin with compactifying 
M theory on $T^7$ with the following brane configuration \cite{tsey}. 
\footnote{There are many other brane configurations that are related 
to this one by U-duality. Three M5 branes intersecting over a line 
with momentum flowing along the line is one such example.}

\bigskip
\rnc{\baselinestretch}{1}
\small\normalsize

\centerline{
\begin{tabular}{cccccccccccc}
Brane & 0 & 1 & 2 & 3 & 4 & 5 & 6 & 7 & 8 & 9 & 10 \cr
M2    & x &   &   &   & x & x &   &   &   &   &    \cr  
M2    & x &   &   &   &   &   & x & x &   &   &    \cr 
M5    & x &   &   &   &   & x &   & x & x & x & x  \cr 
M5    & x &   &   &   & x &   & x &   & x & x & x  \cr 
\end{tabular}
}

\bigskip
\rnc{\baselinestretch}{1.24}
\small\normalsize
\noindent

With suitable choice of the orientation of the branes, 
this configuration breaks $\CN=8$ supersymmetry (SUSY) of the $D=4$ theory
to $\CN=1$. When the number of branes in each group is all equal, 
the background metric describes a direct product of 
an extremal $D=4$ Reissner-Nordstr\"om black-hole and a $T^7$.
See section 3. of \cite{tsey} for more details. 
The $AdS_2\times S^2$ spacetime arising as the near horizon geometry of 
this black-hole is known as the Bertotti-Robinson metric 
\cite{berob}. Note that the brane configuration at hand approaches the 
Bertotti-Robinson metric in the near horizon limit even when the four 
charges are not equal.

The number of SUSY is doubled in the near-horizon limit 
as usual, so we have $D=4$, $\CN=2$ SUSY. 
The super-isometry group of the theory is $SU(1,1|2)$. 
The KK spectrum form representations of the $SU(1,1|2)$ 
superalgebra. 

The methods of the computation used in this paper are well known 
from higher dimensional examples.
There are two approaches to the problem; 
one is direct SUGRA calculation \cite{ads3a}-\cite{ads7a}, 
and the other uses representation theory of superconformal algebra 
together with duality symmetry of SUGRA \cite{ads3b}-\cite{ads7b}. 
We will adopt the first approach and explicitly calculate the spectrum, starting from the $D=11$ SUGRA lagrangian. Although we will be mainly interested in the modes which have bulk degrees of freedom. However, as was noted in ref.\cite{jeremy}, we cannot ignore the boundary modes completely because one of them forms a multiplet with bulk modes. We will make further comments on this point later. 

This paper is organized as follows. 
In section 2, we review the $SU(1,1|2)$ superalgebra and its representation
theory following \cite{ads3b, ads3bb}. 
In section 3, as a warm-up exercise we compute the spectrum of a toy model, 
namely the minimal $D=4$, $\CN=2$ SUGRA. 
This model illustrates many important aspects of the compactification 
on $AdS_2\times S^2$ in a simple setting.
In section 4, we present a summary of our main result.
In section 5 and 6, we sketch the computation of bosonic and fermionic mass
spectrum of the ``realistic'' model obtained from the $D=11$ supergravity . 
We focus on the reduction from $D=11$ to $D=4$ and 
how the $\CN=8$ supermultiplet break into $\CN=2$ multiplets.

\bigskip
As this work was being completed, we received \cite{jeremy}
which has overlap with section 3 of this paper. 
While this paper was being submitted to hep-th e-print archive, 
we received \cite{corley} which considered the same model 
in a manifestly U-duality covariant way.


\section{Review of the $\mathbf{SU(1,1|2)}$ Superconformal Algebra}

The $SU(1,1|2)$ superconformal algebra is defined by the following 
commutation relations
\bea
\slabel{BBcom}
&\left[L_m, L_n \right] = (m-n)L_{m+n}, \  \
\left[J^a, J^b \right] =  i\ep^{abc}J^c, \  \
\left[L_m, J^a \right] =  0,&  \\
\slabel{BFcom}
&\left[L_m, G^{\a\bar{\a}}_r \right] = (\half m - r)G^{\a\bar{\a}}_{m+r}, \  \
\left[J^a, G^{\a\bar{\a}}_r\right] = -\half(\s^a){^\a}_\b G^{\b\bar{\a}}_r,&\\
&\{ G^{\a\bar{\a}}_r, G^{\b\bar{\b}}_s \} = 
\ep^{\bar{\a}\bar{\b}}\{ \ep^{\a\b}L_{r+s} - (r-s)(\s^a\ep)^{\a\b} J^a \}.& 
\slabel{FFcom}
\eea
and the Hermiticity conditions
\be
L_m^\dagger = L_{-m}, \  \ (J^a)^\dagger = J^a, \  \
(G^{\a\bar{\a}}_r)^\dagger = 
\ep_{\a\b}\ep_{\bar{\a}\bar{\b}} G^{\b\bar{\b}}_{-r}.
\label{hermi}
\ee
The bosonic generators $L_{+1,0,-1}$ and $J^{0,1,2}$ generate 
the $SL(2,\IR)$ conformal group 
and the $SU(2)$ $R$-symmetry group, respectively. 
We have eight supercharges all together; 
$G_{\pm 1/2}^{\a\bar{\a}}$ carry $L_0$ charge $\mp \half$ and 
transform in (${\mathbf \half},{\mathbf \half})$ representation of 
$SU(2)_R\times SU(2)_{Aut}$, where the second $SU(2)$ is a global 
automorphism.

One explicit way to find the representations of a superalgebra 
is the oscillator construction \cite{osc1, osc2, osc3}. 
The oscillator representation of the generators of $SU(1,1|2)$ 
is given by 
\bea
\slabel{losc}
L_{-1} = \vec{a}^{\dagger}_1\cdot\vec{a}^{\dagger}_2,
&L_0 =\half(\vec{a}_1\cdot\vec{a}^{\dagger}_1 + 
\vec{a}^{\dagger}_2\cdot\vec{a}_2), & 
L_{+1} = \vec{a}_2\cdot\vec{a}_1 \\
J^+ = \vec{\psi}^{\dagger}_1\cdot \vec{\psi}^{\dagger}_2,
&J^0 = \half(\vec{\psi}^{\dagger}_1\cdot \vec{\psi}_1 -
\vec{\psi}_2\cdot \vec{\psi}^{\dagger}_2), & 
J^- =  \vec{\psi}_2\cdot \vec{\psi}_1
\slabel{josc} 
\eea
\be
G_{-1/2} = \left[\matrix{ 
G^{--}_{-1/2} & G^{-+}_{-1/2} \cr G^{+-}_{-1/2} & G^{++}_{-1/2}
}\right]
= \left[ 
\matrix{
\vec{a}^{\dagger}_2 \cdot \vec{\psi}_1 &
-\vec{a}^{\dagger}_2\cdot \vec{\psi}_2 \cr
\vec{a}^{\dagger}_2 \cdot \vec{\psi}^{\dagger}_2 &
\vec{a}^{\dagger}_1 \cdot \vec{\psi}^{\dagger}_1
}
\right],\  \ 
G_{+1/2} = \left[
\matrix{
\vec{a}_1 \cdot \vec{\psi}_1 &
-\vec{a}_2\cdot \vec{\psi}_2 \cr
\vec{a}_1 \cdot \vec{\psi}^{\dagger}_2 &
\vec{a}_2 \cdot \vec{\psi}^{\dagger}_1 
}
\right],
\ee
where $\vec{a}^{\dagger}_i, \vec{a}_i$ are $n$-component vectors of 
bosonic creation and annihilation operators, and 
$\vec{\psi}^{\dagger}_i, \vec{\psi}_i$ are the fermionic counterparts.
It is straightforward to see that they satisfy all the commutation relations 
and Hermiticity conditions \eq{BBcom}-\eq{hermi} except \eq{FFcom},
which is modified as
\bea
\{ G^{\a\bar{\a}}_r, G^{\b\bar{\b}}_s \} &=& \ep^{\bar{\a}\bar{\b}}
\{ \ep^{\a\b}L_{r+s} - (r-s)(\s^a\ep)^{\a\b} J^a + \ep^{\a\b} I \}, \\
I &\equiv& 
\half(\vec{a}^{\dagger}_1\cdot\vec{a} - \vec{a}^{\dagger}_2\cdot\vec{a}_2)
-\half(\vec{\psi}^{\dagger}_1\cdot \vec{\psi}_1 -
\vec{\psi}^{\dagger}_2\cdot \vec{\psi}_2).
\eea
The extra $U(1)$ generator $I$ must be added in order for the algebra 
to be closed. 
However, since $I$ commutes with all the other generators, 
we may work in the restricted Fock space on which $I=0$, 
where the algebra precisely reduces to that of $SU(1,1|2)$.
\footnote{We thank Jan de Boer for a correspondence on this point.}

For a given integer $n$, 
the oscillator vacuum is identified with the lowest $J^0$-weight state 
of a chiral primary operator. We act $G_{-1/2}^{+\pm}$ on the vacuum to obtain the 
lowest weight states of other primary operators in the supermultiplet. 
Higher weight states of a given operator are obtained by acting $J^+$ 
on the lowest weight state. 

The quantum numbers of each state is easily computed using the explicit 
oscillator representation of the generators \eq{losc}, \eq{josc}. 
The quantum numbers of the lowest weight state of each primary operator
are summarized in Table \ref{superalg}. The total angular momentum  $j$ 
is defined by $\vec{J}^2 =j(j+1)$. 
The number of states for each primary operator, $2j+1$, 
is also included in the table. 
\mytable{superalg}
{
\begin{tabular}{r|cccc} \hline
Lowest weight states & $j$ & $L_0$ & Degeneracy \cr \hline
$|0\rangle$ & $n/2$ & $n/2$ & $n+1$ \cr
$G_{-1/2}^{++}|0\rangle, G_{-1/2}^{+-}|0\rangle$ 
& $(n-1)/2$ & $(n+1)/2$ & $2\times n$ \cr 
$G_{-1/2}^{++}G_{-1/2}^{+-}|0\rangle$  
& $(n-2)/2$ & $(n+2)/2$ & $n-1$ \cr \hline
\end{tabular}
}
{The short multiplets of $SU(1,1|2)$ superconformal algebra labeled by 
an integer $n$} 

\section{Toy Model \label{toysec}}

As a warm-up exercise, we compute the mass spectrum of the minimal 
$D=4$, $\CN=2$ SUGRA. It is the simplest SUGRA 
that admits the $AdS_2\times S^2$ solution with the $SU(1,1|2)$ 
superalgebra. The theory contains a single $\CN=2$ gravity multiplet 
whose component fields are a graviton, a massless vector and 
a complex gravitino. 

\subsection{Result}

The computation in the following subsections show that the KK spectrum of 
the toy model contains the short multiplets in Table \ref{superalg} 
for all even $n$.
We have two copies of each multiplet for $n\ge 4$ and one copy for $n=2$. 
The result is depicted in Figure \ref{toy}.

\fig{360pt}{toy}{\small
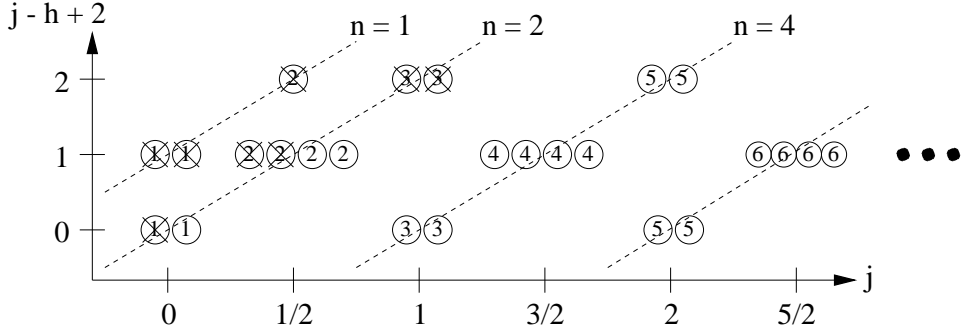
The complete KK spectrum of the toy model. 
Each circle in the figure represents a state which has a definite value of $h$
and $j$. The crossed circles correspond to the boundary states. The degeneracy ($2j+1$) of each state is included in the circle. 
The states belonging to the same $SU(1,1|2)$ multiplet is connected by a 
dotted line. The two KK towers on the top row satisfy $h=j$ and correspond to 
chiral primary states.
}

{}From the point of view of the SUGRA computation, each physical degree of 
freedom of the fields in $D=4$ give a KK tower. That explains why we have
four bosonic and four fermionic series of states in the spectrum. 
Depending on the spin and the polarization of a given field, 
the low lying modes ($j=0,\half,1$) modes may be absent. 
Some of other low lying modes become massless and can be gauged away 
from the bulk spectrum. 
The absence of such modes is necessary in order for the KK spectrum 
to arrange itself into representations of $SU(1,1|2)$.

In addition to the bulk degrees of freedom, there may be modes that are pure 
gauge in the bulk but can live on the boundary. The authors of 
\cite{jeremy} showed that the boundary modes indeed exist and form one $n=2$ 
and one $n=1$ representations of $SU(1,1|2)$ algebra. 
We included these boundary modes in Figure \ref{toy} for completeness.
In particular, we cannot ignore them since one of them forms $n=2$ multiplet with bulk modes, as can be seen from the figure. 
\subsection{Bosonic mass spectrum}

\subsubsection{Setup}

We normalize the fields such that the action reads
\be
2\kappa^2 S = \int d^4 x \sqrt{-G} 
\left\{ 
R - \recip{ 4} F^2  -\bar{\psi}_m \G^{mnp}\del_n\psi_p 
-\textstyle{i\over 2}\bar{\psi}_m(F^{mn} + {\half}F_{rs}\G^{rsmn})\psi_n 
\right\},
\ee
up to terms quartic in fermions that are irrelevant to our computation.
 The bosonic equations of motion consist of the Einstein and Maxwell equations
in vacuum.
\be
R_{mn} = \half F_{ml} {F_n}^l - \recip{ 8} G_{mn} F^2, \  \
\del^m F_{mn} = 0.
\ee
These equations admit dyonic Reissner-Nordstr\"om black-hole solutions. 
The near horizon geometry of an extremal black-hole gives the
$AdS_2\times S^2$ solution. The radius of the $S^2$ is equal to the
Schwarzchild radius of the black-hole.
For simplicity, we consider an extremal electric black-hole with unit radius 
only. Then the $AdS_2\times S^2$ solution reads
\be
\begin{array}{ll}
R_{\m\n\l\s} = - (g_{\m\l}g_{\n\s} -g_{\m\s}g_{\n\l}), & 
\bar{F}_{\m\n} = 2 \ep_{\m\n},  \cr
R_{\a\b\ga\d} = g_{\a\ga}g_{\b\d} - g_{\a\d}g_{\b\ga}, &  
\bar{F}_{\a\b} = 0. \cr
\end{array}
\ee
where the Greek letters $\a, \b \cdots$ and $\m \n \cdots$ label two dimensional indices for $AdS_2$ and $S^2$ respectively. 
The fermions are set to zero.
We are interested in the mass spectrum of the fluctuations of 
the fields around this background. We use the following parametrizations of
the fluctuations.
\bea
G_{\a\b} = g_{\a\b} + h_{(\a\b)} + \half h_2 g_{\a\b}, \  \
&G_{\m\n} = g_{\m\n} + h_{(\m\n)} + \half h_1 g_{\m\n},& \  \
G_{\m\a} = h_{\m\a}, \\
F_{\a\b} = \del_\a a_\b - \del_\b a_\a, \  \
&F_{\m\n}= 2\ep_{\m\n} + \del_\m a_\n - \del_\n a_\m,& 
\eea
where the paranthesis denotes the traceless part of a symmetric tensor.

\subsubsection{Spherical harmonics decomposition and gauge choice}

Each field can be decomposed into spherical harmonics. Unlike higher 
dimensional spheres, $S^2$ does not have genuine
vector or tensor spherical harmonics.
Vector and tensor fields are spanned by derivatives of the scalar spherical
harmonics. 
\be
\begin{array}{ll}
h_{(\a\b)} = \phi_1^I \del_{(\a}\del_{\b)} Y^I 
+ \phi_2^I \ep_{\ga(\a}\del_{\b)} \del^\ga Y^I\  \ (j\ge 2), &
h_2 = h_2^I Y^I ,\cr
h_{(\m\n)} = h_{(\m\n)}^I Y^I, &
h_1 = h_1^I Y^I, \cr
h_{\m\a} = w_\m^I \del_\a Y^I + v_\m^I \ep_{\a\b} \del^\b Y^I\  \ (j\ge 1),&\cr
a_\a = q^I \del_\a Y^I + b^I \ep_{\a\b} \del^\b Y^I \  \ (j\ge 1), &
a_\m = a_\m^I Y^I. \cr
\end{array}
\ee
The composite index $I$ specifies both the total angular momentum
$j$ and the $J^0$ eigenvalue $m$.
The restrictions on the value of $j$ for some fields are due to the fact that
\be
\del_\a Y^{(j=0)} = \del_{(\a}\del_{\b)} Y^{(j=1)} = 
\ep_{\ga(\a}\del_{\b)} \del^\ga Y^{(j=1)} = 0.
\ee
Not all the modes in the above expansion are physical since the
($D=4$) graviton and gauge fields are subject to the 
gauge transformations,
\bea
\d h_{mn} &=& \del_m \Lambda_n + \del_n \Lambda_m, \\
\d a_m &=& -\bar{F}_{mn}\Lambda^n + \del_m (\Sigma + \Lambda^n \bar{A}_n).
\eea
The functions $\Lambda_m$ and $\Sigma$ are also expanded 
in spherical harmonics. 

We need to make a choice of gauge. 
First, consider the case $j\ge 2$.
We can gauge away $h_{(\a\b)}$ completely by 
a suitable choice of $\Lambda_\a$. 
We then use $\Lambda_\m$ to eliminate the $w_\mu^I$ terms. 
Lastly, we use $\Sigma$ to eliminate the $q^I$ terms. 
With this choice of gauge, we note that
\be
\del^\a h_{\m\a} = 0,\  \ \del^\a a_\a = 0.
\ee
For $j=1$, $h_{(\a\b)}$ modes are absent, so $\Lambda_\a$ can be used 
to reduce other degrees of freedom. We find it convenient to
parametrize $\Lambda_m$ and $\Sigma$ as
\bea
\Lambda_\m^{(1)} &=& (K_\m + \del_\m X) \cdot Y, \\
\slabel{resga}
\Lambda_\a^{(1)} &=& P \cdot \ep_{\a\b}\del^\b Y - X\cdot \del_\a Y, \\
\Sigma^{(1)} &=& Q\cdot Y,
\eea
where the dot product means the sum over the three components of 
$j=1$ spherical harmonics.
We can use $X$, $K_\m$ and $Q$ to gauge away $h_1$, $w_\m$ and $q$, 
respectively. 
Under the gauge transformation by $P$, $v_\m$ is shifted by 
$\del_\m P$. 
This indicates that $v_\m$ is a massless gauge field in $AdS_2$.
Indeed, the mass term for $v_\m$ is absent as we will see below.
Being a gauge field in $D=2$, $v_\m$ has 
no propagating degree of freedom in the bulk. Also $h_2$ can be locally gauged away by residual gauge symmetry.

For $j=0$, $h_{(\m\n)}$, $h_1$, $h_2$ and $a_\m$ are 
the only modes that remain. 
The gauge parameter $\Lambda_\a$ is absent.
We can use $\Lambda_\m$ to gauge away $h_{(\m\n)}$. 
The vector $a_\m$ becomes a gauge field in $AdS_2$ 
with $\Sigma$ being the gauge transformation parameter, 
and again has no bulk degree of freedom.

\subsubsection{Linearized field equations}

The linearized Einstein and Maxwell equations read
\be
\begin{array}{rcl}
R^{(1)}_{mn}(h) &=&
\bar{F}_n^{\;\;l} (\del_m a_l - \del_l a_m)
+ \bar{F}_m^{\;\;l} (\del_n a_l - \del_l a_n)
-g_{mn} \bar{F}^{kl} \del_k a_l \cr
&& - \bar{F}_{mk} \bar{F}_{nl} h^{kl} 
+ \half g_{mn} \bar{F}_j^{\;\;k} \bar{F}^{jl} h_{kl} 
-\recip{ 4} h_{mn} \bar{F}^2,
\end{array}
\ee
\be
\del^m(\del_m a_n - \del_n a_m) 
- \half (2\del^m h_{ml} - \del_l h) \bar{F}^l_{\;\;n} 
- \del_m h_{ln} \bar{F}^{ml} = 0,
\ee
where the linearized Ricci tensor is defined by
\be
R^{(1)}_{mn}(h) \equiv -\del^2 h_{mn} - \del_m \del_n h^k_k 
+ \del^k\del_m h_{nk} + \del^k\del_n h_{mk}.
\ee
Upon spherical harmonics decomposition, the $\a\b$ component of the Einstein 
equation yields the following three equations. They are the coefficients of
$g_{\a\b} Y^I$, $\del_{(\a}\del_{\b)}Y^I$, and 
$\ep_{\ga(\a}\del_{\b)}\del^\ga Y^I$, respectively. 
\bea
\slabel{tb1}
\del_x^2 h_2^I -j(j+1)(h_1^I +h_2^I) 
&=& 4\ep^{\m\n}\del_\m a_\n^I -2 h_2^I + 4h_1^I, \\
\slabel{tb2}
h_1^I &=& 0, \\
\slabel{tb3}
\del^\m v_\m^I &=& 0.
\eea
We separate the trace and traceless part of the $\m\n$ component of 
the Einstein equation. We replace $h_{\m\n}$ by $h_{(\m\n)}$ in all the 
equations below using the constraint \eq{tb2}.
\bea
\slabel{tb4}
\del_x^2 h_2^I -2\del^\m\del^\n h_{(\m\n)}^I 
&=& -4\ep^{\m\n}\del_\m a_\n^I, \\
\slabel{tb5}
\del_x^2 h_{(\m\n)}^I -j(j+1)h_{(\m\n)}^I +2 h_{(\m\n)}^I  
&=& \del_\m\del^\l h_{(\l\n)}^I +\del_\n\del^\l h_{(\l\m)}^I
-g_{\m\n} \del^\l \del^\s h_{(\l\s)}^I \xx
&& -\del_{(\m}\del_{\n)} h_2^I 
\eea
The $\m\a$ component of the Einstein equation splits into two pieces. 
They are the coefficients of $\del_\a Y^I$ and $\ep_{\a\b}\del^\b Y^I$,
respectively.
\bea
\slabel{tb6}
\del_\m h_2^I - 2 \del^\n h_{(\n\m)}^I &=& - 4 {\ep_\m}^\n a_\n^I,  \\
\del_x^2 v_\m^I - (j^2+j-3) v_\m^I &=& 2\ep_{\m\n}\del^\n b^I.
\slabel{tb7}
\eea
The $\a$ component of the Maxwell equation splits in the same way,
\bea
\slabel{tb8}
\del^\m a_\m^I &=& 0,  \\
\del_x^2 b^I - j(j+1) b^I &=& 2\ep^{\m\n} \del_\m v_\n^I.
\slabel{tb9}
\eea
The $\m$ component of the Maxwell equation yields a single equation,
\be
\label{tb10}
\del_x^2 a_\m^I - (j^2+j-1)a_\m^I = \ep_{\m\n} \del^\n h_2^I.
\ee

\subsubsection{Computation of the mass spectrum: $\mathbf{j\ge 2}$}

Altogether, we have ten equations of motion \eq{tb1} - \eq{tb10}. 
We already used \eq{tb2} to eliminate $h_1^I$. 
We also note that \eq{tb6} implies \eq{tb4} for $j\ge1$.
So, the number of independent equations is eight.
We put off the discussion of \eq{tb5} and \eq{tb6} to the end of 
this subsection.

Among the other six equations, \eq{tb3} and \eq{tb8} are constraints, 
and the other four are dynamical equations for each physical field.
We first use the constraints to set on shell.
\footnote{This type of transformations appear in other compactifications 
with electric background field strength. For example, see \cite{ads4a}.} 
\be
\label{onshdual}
v_\m = 2\ep_{\m\n} \del^\n v, \  \ a_\m = \ep_{\m\n} \del^\n a
\ee
To simplify notations, we are suppressing the superscripts $I$ 
in the equations from here to the end of this subsection.
Inserting these in \eq{tb1} and \eq{tb9} immediately yields
\bea
\slabel{tb11}
\del_x^2 h_2 - (j^2+j-2)h_2 - 4\del_x^2 a &=& 0, \\
\del_x^2 b - j(j+1) b - 4 \del_x^2 v  &=& 0.
\slabel{tb12}
\eea
After some manipulations, the other two equations \eq{tb7} and \eq{tb10} give
\bea
\slabel{tb13}
\del_x^2 v - (j^2+j-2) v - b &=& 0, \\
\del_x^2 a - j(j+1) a - h_2  &=& 0.
\slabel{tb14}
\eea
They are diagonalized by the following linear combinations of the fields.
\bea
s_1 &=& b-2(j+2)v,\  \ s_2 = h_2-2(j+1)a, \\
t_1 &=& b+2(j-1)v,\  \ t_2 = h_2 + 2ja.
\eea
They satisfy
\bea
\del^2 s_i - j(j-1)s_i &=& 0, \\
\del^2 t_i - (j+1)(j+2)t_i &=& 0. 
\eea

In $AdS_2$, the scaling dimension of the operator corresponding to 
a scalar field is given by \cite{gkp, witt} 
\be
h = \half(1 + \sqrt{1+ 4 m^2}).
\ee
This implies that the fields $s_{1,2}$ have $h=j$ and are chiral primaries, 
while $t_{1,2}$ have $h=j+2$. 

It remains to analyze \eq{tb5} and \eq{tb6}. Inserting \eq{tb6} into \eq{tb5}
and using \eq{onshdual}, we find
\be
\del_x^2 h_{(\m\n)}^I -j(j+1)h_{(\m\n)}^I +2 h_{(\m\n)}^I  
= 4\del_{(\m}\del_{\n)} a.
\ee
It is also possible to show that {\em in two dimensions}, \eq{tb6} implies
\be
(\del^2 + 2) h_{(\m\n)} = \del_{(\m}\del_{\n)} (h_2 +4a)
\ee
It can be derived most easily in a light-cone coordinate and a conformal gauge.
Combining these two equations, we find that
$h_{(\m\n)}$ is algebraically determined by $h_2$ and hence has 
no degree of freedom. This argument is valid for $j=1$ also, 
but not for $j=0$.

\subsubsection{$\mathbf{j=1}$}

The computation for $j=1$ differs from that for $j\ge 2$ in two ways.
First, $h_1$ is removed by a gauge choice rather than 
the constraint \eq{tb2} which is absent because 
$\del_{(\a}\del_{\b)}Y^{(j=1)}=0$. 
Second, $v_\m$ and $h_2$ has no bulk degree of freedom and can be eliminated.  %
The equations can be diagonalized as before, 
and the three eigenstates are identified with the
$j=1$ points of the $t_1$, $t_2$ and $s_2$ series. 
The absence of the corresponding point on the $s_1$ series 
is a consequence of the fact that $v_\m$ is a gauge field.
Note also that $s_2$ can be gauged away {\em on shell} by 
a residual gauge degree of freedom. By $X$ in \eq{resga} with $\del^2 X =0$, 
$s_2$ is shifted to $s_2 + X$. Therefore it is also a boundary degrees of freedom.

\subsubsection{$\mathbf{j=0}$}

We have the field equations for $h_1$, $h_2$ and $a_\m$
\bea
\del_x^2 (h_1 + h_2) &=& -4\ep^{\m\n}\del_\m a_\n -2 h_1, \\
\del_x^2 h_2 &=& 4\ep^{\m\n}\del_\m a_\n -2h_2 + 4h_1, \\
\del^\n(\del_\n a_\m - \del_\m a_\n) &=& \ep_{\m\n}\del^\n (h_2 - h_1).
\eea
Recall that we gauged away $h_{(\m\n)}$. Its equation of motion then 
gives a ``Gauss law'' constraint,
\be
\del_{(\m}\del_{\n)} h_2 = 0.
\ee
One can easily show that (in light cone coordinate, for example) 
the only normalizable solution to the constraint is $h_2 = $ constant.
It is consistent to set $h_2$ to zero.
We can eliminate the gauge field $a_\m$ from the $h_1$ equation and find that 
$m^2 = 2$. This is identified with the $j=0$ point of the $t_2$ series.
This completes the derivation of the bosonic spectrum in Figure \ref{toy}.

\subsection{Fermionic mass spectrum}

The linearized field equation for the fermion reads
\be
\G^{mnp}\del_n\psi_p = -\textstyle{i\over 2}
\left(\bar{F}^{mn}+\half \bar{F}_{rs}\G^{mnrs}\right)\psi_n.
\label{toyferm}
\ee
The linearized SUSY transformation law plays the role of a gauge symmetry,
that is, the following variation leaves the field equation invariant:
\be
\d \psi_m = \del_m \ep - i\bar{F}_{nl}
\left(\recip{ 4}\G^l\d^n_m -\recip{ 8} \G{_m}^{nl}\right) \ep.
\ee
It is convenient to separate the ``trace'' and the ``traceless'' part of 
$\psi_\m$ and $\psi_\a$.
\be
\psi_\m = \psi_{(\m)} + \half\G_\m \l,\  \ 
\psi_\a = \psi_{(\a)} + \half\G_\a \eta\  \
(\G^\m\psi_{(\m)} = \G^\a\psi_{(\a)} = 0).
\ee
We decompose the $D=4$ gamma matrices in terms of the $D=2$ gamma matrices
as follows:
\be
\G^\m = \ga^\m \otimes \identity,\  \ 
\G^\a = \bar{\ga} \otimes \tau^\a \  \ (\bar{\ga} = \ga^0\ga^1).
\ee
Now we can split \eq{toyferm} into four components
\bea
\slabel{chieom}
(\dirac_{x}+\bar{\ga}\dirac_{y})\eta + \bar{\ga}\dirac_y\l 
-2\del^\a\psi_{(\a)} &=& -i\bar{\ga}\l, \\
\slabel{lameom}
(\dirac_{x}+\bar{\ga}\dirac_{y})\l + \dirac_x\eta -2\del^\m\psi_{(\m)} &=&
-i\bar{\ga}\eta, \\
\slabel{psimeom}
\del_{(\m)}\eta + \bar{\ga}\dirac_y\psi_{(\m)} &=& -i\bar{\ga}\psi_{(\m)}, \\
\slabel{psiaeom}
\del_{(\a)}\l + \dirac_x\psi_{(\a)} &=& +i\bar{\ga}\psi_{(\a)},
\eea
where the first two equations are the traceless parts of the $\m$ and $\a$
components of \eq{toyferm}, respectively. The other two are the trace parts. Here, $\dirac_x, \dirac_y$  are the two dimensional dirac operators.
The gauge transformation law also divides into four pieces.
\bea
\d\psi_{(\m)} &=& \del_{(\m)}\ep,\  \ \d\l = \dirac_x\ep - i\bar{\ga}\ep, \\
\d\psi_{(\a)} &=& \del_{(\a)}\ep,\  \ \d\eta = \bar{\ga}(\dirac_y\ep + i\ep). 
\eea
Consider the spherical harmonics decomposition.
\be
\begin{array}{ll}
\l = \l_+^I \Sigma_+^I + \l_-^I \Sigma_-^I, &
\psi_{(\m)} = \psi_{(\m)+}^I \Sigma_+^I + \psi_{(\m)-}^I \Sigma_-^I, \\
\eta = \eta_+^I \Sigma_+^I + \eta_-^I \Sigma_-^I, &
\psi_{(\a)} = \psi_+^I\del_{(\a)} \Sigma_+^I +\psi_-^I \del_{(\a)}\Sigma_-^I,\\
\ep =  \ep_+^I \Sigma_+^I + \ep_-^I \Sigma_-^I. &
\end{array}
\ee
See Appendix \ref{sphh} for the definition and properties of 
spinor spherical harmonics.
For $j\ge 3/2$, it is clear that one can gauge away $\eta$ completely. 
Then \eq{psimeom} sets $\psi_{(\m)}^I = 0$. 
In turn, we find in \eq{lameom} that $\l$ satisfies
the eom for a free massless spinor in $d=4$.
Finally, \eq{chieom} determines $\psi_{(\a)}$ algebraically 
in terms of $\l$. As a consistency check, we substitute it into \eq{psiaeom} 
and find the same eom for $\l$.
Thus all that remains is to find the mass spectrum of $\l$.
After the spherical harmonics decomposition, the equation reduces to
\be
\dirac \l_+ + i(j+\half)\bar{\ga}\l_+ = 0,\  \ 
\dirac \l_- - i(j+\half)\bar{\ga}\l_- = 0. 
\ee
The mass eigenstates are $\xi_1 = (1+i\bar{\ga})E$ and 
$\xi_2 = (1-i\bar{\ga})F$ both of which have $m = j+\half$.
\footnote{
We may choose $(1-i\bar{\ga})E$ and $(1+i\bar{\ga})F$. 
The two choices are not independent, since one can multiply either of them 
by $\bar{\ga}$ to get the other.
}

The computation is slightly different for $j=1/2$.
To begin with, 
we note the following property of the $j=1/2$ spherical harmonics.
\be
\del_\a \Sigma_{\pm} = \pm \textstyle{i\over 2} \t_\a \Sigma_{\pm} \  \
\imp \  \ 
\dirac_y \Sigma_{\pm} = \pm i \Sigma_{\pm}.
\ee
It has three consequences. 
First, modes for $\psi_{(\a)}$ are absent.
Second, \eq{psiaeom} is trivially satisfied. 
Finally, the gauge variation of $\eta_-$ vanishes for arbitrary $\ep_-$.
We choose to gauge away $\eta_+$ and $\psi_{(\m)-}$ 
using $\ep_+$ and $\ep_-$, respectively.
With these in mind, we analyze the three field equations. 
{}From the coefficients of $\Sigma_+$ in \eq{chieom} and \eq{psimeom}, 
we find that
\be
\l_+=0,\  \ \psi_{(\m)+} = 0.
\ee
The coefficients of $\Sigma_-$ of the same equations yield
\be
(\dirac_x - i\bar{\ga})\eta_- =0,\  \ \del_{(\m)}\eta_- = 0.
\ee
These two equations together imply that $\eta_-$ has 
no propagating degree of freedom and can be set to zero consistently.
Finally \eq{lameom} gives
\be
(\dirac_x - i\bar{\ga})\l_- = 0,
\ee
which we recognize as the $j=1/2$ point of the $\xi_2$ series.

The scaling dimension of the operator corresponding to 
a spinor field in $AdS_2$ is given by 
\be
h = |m|+\half,
\ee
which implies that the fields $\xi_{1,2}$ have $h=j+1$.

\section{Summary of the Main Result}

\fig{360pt}{real}
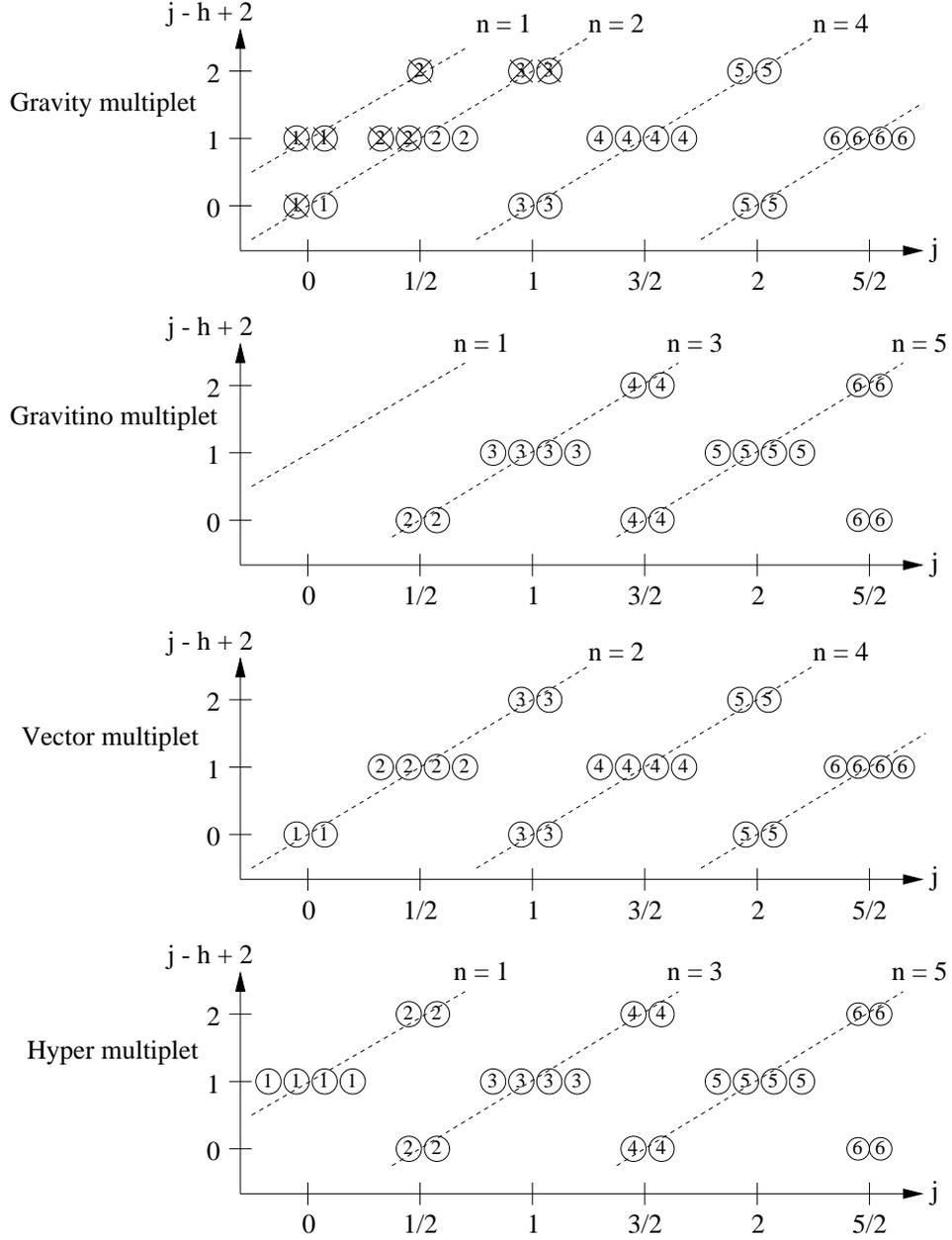
{The KK spectrum of the $D=11$ SUGRA on $AdS_2\times S^2\times T^7$}

We now turn to the model which is the main interest of this paper, namely, 
the one obtained from the low energy M theory. 
We first dimensionally reduce $D=11$ SUGRA to obtain $D=4$. 
The resulting $\CN=8$ SUGRA contains 
1 graviton, 8 real gravitini, 28 vectors, 56 real spinors and 70 scalars.
Compactification on $AdS_\times S^2$ keeps $\CN=2$ SUSY unbroken. 
In the $\CN=2$ language, we have 1 gravity, 6 gravitino, 15 vector and 
10 (complex) hyper multiplets. Each multiplet has 4 bosonic and  4 fermionic real degrees 
of freedom. The $4+4$  KK towers arrange themselves into representations 
of $SU(1,1+2)$ superalgebra. 
Figure \ref{real} describes the KK spectrum of each multiplet. 
The gravity multiplet is identical to that of the toy model. 
The vector multiplet is similar to the gravity multiplet, but it has
two copies of the $n=2$ representation. 
Gravitino multiplet contains two copies of representations for all odd 
$n$ except for $n=1$.
Hyper multiplet includes the $n=1$ representation.

The analysis for the boundary modes is more complicated since one has to keep track of modes which may be removed by fixing gauges. We will concentrate on obtaining bulk modes. As in the toy model, we included the boundary degrees of freedom for 
the gravity multiplet in the figure. 
Boundary degrees of freedom can arise in the gravitino multiplet as well, 
but are not determined by the computation here. 

In the following two sections, we explain the dimensional reduction of 
the field equations from $D=11$ to $D=4$, how different fields fall into 
$\CN=2$ multiplets and how each multiplet produces the KK spectrum given 
in Figure \ref{real}.


\section{Bosonic Mass Spectrum}

\subsection{Setup}

We normalize the fields such that the action reads
\be
\begin{array}{rcl}
2\kappa^2 S &=& \int d^{11} x \sqrt{-G} 
\left\{ R - \recip{ 2\cdot 4!} F^2 -\bar{\Psi}_I \G^{IJK}\del_J\Psi_K \right\} 
+ \recip{ 3!} \int A\wedge F \wedge F \cr
&&+\int d^{11} x \sqrt{-G} 
\left\{
\recip{ 4\cdot 4!} \bar{\Psi}_I (\G^{IJKLMN} \Psi_J F_{KLMN}
+12 \G^{KL}G^{MJ} F{^I}_{KLM}) \Psi_J 
\right\}.
\end{array}
\ee
The terms quartic in $\Psi_M$ are not relevant to this paper and have been
omitted. 
Bosonic equations of motion consist of the Einstein and Maxwell equations
in vacuum.
\bea
R_{MN} &=& \recip{ 2\cdot 3!} F_{MIJK} {F_N}^{IJK} - 
\recip{ 6\cdot 4!} G_{MN} F^2, \\
\del^M F_{MIJK} &=& 
\recip{ 2\cdot 4!\cdot 4!} \ep_{IJKL_1L_2L_3L_4M_1M_2M_3M_4} 
F^{L_1L_2L_3L_4}F^{M_1M_2M_3M_4}.
\eea
The $AdS_2\times S^2\times T^7$ solution is given by
\be
\begin{array}{rcl}
ds_{11}^2 &=& g_{\m\n}dx^\m dx^\n + g_{\a\b} dx^\a dx^\b 
+ \delta_{ab} dz^a dz^b +\delta_{st} dw^s dw^t, \\
R_{\m\n\l\s} &=& - (g_{\m\l}g_{\n\s} - g_{\m\s}g_{\n\l}),\  \  
R_{\a\b\ga\d} = g_{\a\ga}g_{\b\d} - g_{\a\d}g_{\b\ga}, \\
\bar{F}_{\m\n 45} &=& \bar{F}_{\m\n 67} = \ep_{\m\n}, \  \ 
\bar{F}_{\a\b 46} = \bar{F}_{\a\b 75} = \ep_{\a\b}.
\end{array}
\label{bg}
\ee
All other fields are set to zero. This is the near horizon geometry of 
the brane configuration shown in the introduction. The notational conventions for the indices are summarized in Appendix A.

Note that the background fields are self-dual in the $SO(4)$ 
for the four coordinates along which the branes lie. 
Equivalently, it transforms in the $(\mathbf{0},\mathbf{1})$ representation of
$SO(4) \simeq SU(2)_+ \times SU(2)_-$.
This follows from the requirement of partially unbroken supersymmetry.
If any one of the relative sign for the gauge field is flipped, 
the supersymmetry is completely broken, even though it is still a solution 
to the equations of motion.  
\footnote{In fact, supersymmetry requres that the product of the signs of the
gauge field be $+1$. We set all the signs to be $+1$ using coordinate 
redefinition and parity transformation.}  
This background breaks the $SO(7)$ isometry of the internal $T^7$ down to 
$SU(2)_+\times SU(2)_3$, where $SU(2)_3$ is the rotation of $w^{8,9,10}$. 
These internal symmetries will play a crucial role in grouping the fields, 
as will be shown in the next section.  

\subsection{Linearized field equations and reduction to $\mathbf{D=4}$}

We linearize the equations in $D=11$ in fluctuations around the background,
\be
G_{MN} = g_{MN} + h_{MN}, \  \
F_{IJKL} = \bar{F}_{IJKL} + 4 \del_{[I} a_{JKL]},
\ee
and then dimensionally reduce it to $D=4$ by keeping only the zero modes of 
the fluctuations in internal $T^7 = T^4 \times T^3$. 
We then redefine some of the fluctuation fields,
\be
\begin{array}{ll}
h^{(11)}_{mn} = h^{(4)}_{mn} -\half (B^a_a + B^s_s) g_{mn},\  \ 
h_{ab} = B^{ab}, \  \
h_{ma} = V^a_m, \\
a_{abc} = C^{abc}, \  \ 
a_{mab} = A^{ab}_m, \  \ 
a_{mna} = D^a_{mn}, &\\
\end{array}
\ee
The definitions of $B, V, C, A, D$ remains valid when we replace 
the $a,b$ indices by $s,t$ indices.
The shift in $h_{mn}$ is the linearized version of the Weyl rescaling 
which is necessary to absorb the volume factor of the internal dimensions 
and put the action into the standard Einstein-Hilbert form.

Also, one can do Hodge dual transformation to reduce the indices of 
the tensor fields.
The tensor field with three index, $a_{mnl}$ is the most trivial one, 
its dual field having rank $-1$ formally. This implies it has no dynamics. 
Indeed, one can show explicitly from its equations of motion that
it has no degree of freedom and decouples from all the other fields. 
The next one we consider is the rank two tensor field $D^a_{mn}$ whose 
linearized equation of motion is
\be
\nabla^m \{ \del_{[m} D^a_{ij]} + \bar{F}^{ab}_{[ij}V^b_{m]} \} = 0
\ee
which turns into an identity if we introduce the dual scalar
\be
3 \nabla^{[l}D^{mn] a} + 3 \bar{F}^{ab}_{[ij}V^b_{m]}
= \epsilon^{lmnk} \nabla_k D^a
\ee
Then the Bianchi identity for the original $D^{lmn a}$ turns into 
the equation of motion for $D^a$,
\be
\del^2 D^a = \recip{ 4} \ep^{klmn}\bar{F}^{ab}_{kl}W^b_{mn},
\ee
where $W^a_{mn}$ is the field strength of $V^a_{mn}$.
The equation remains valid when $a$ is replaced by $s$, 
except that in this case the right-hand side vanishes.

The quantum numbers of the various fluctuation fields with respect to 
the internal symmetries are summarized below, along with that of the 
background gauge field $\bar F^{ab}_{mn}$.  Using this table, 
one can divide the fields into small groups, 
where the fields belonging to the same group can couple to each other. 
The fields within a group must have the same quantum numbers except 
the broken $SU(2)_-$ charge, which can be shifted by $1$ by the background 
field. We label these groups by capital roman letters.  
Note that we separated the self-dual and the anti-self-dual parts 
of the fields which are rank two tensors in $SO(4)$ by
\be
A^{ab\pm} \equiv {1 \over 2}(A^{ab} \pm \epsilon^{abcd} A_{cd}).
\ee

\mytable{bosongrp}
{
\begin{tabular}{c|cc|c|c}
\hline
Field 	        &$SU(2)_+$&$SU(2)_-$&$SU(2)_3$& Group \cr \hline\hline
$h_{mn}$	&  0  &  0  & 0 & F \cr \hline
$V_m^a$		& 1/2 & 1/2 & 0 & D \cr
$V_m^s$		&  0  &  0  & 1 & B \cr \hline
$B_{(ab)}$	&  1  &  1  & 0 & E \cr
$3B^a_a+2B^s_s$	&  0  &  0  & 0 & A \cr
$B_{as}$	& 1/2 & 1/2 & 1 & C \cr
$B_{(st)}$	&  0  &  0  & 2 & A \cr
$B^s_S$		&  0  &  0  & 0 & F \cr \hline\hline
$A_m^{ab+}$	&  0  &  1  & 0 & F \cr
$A_m^{ab-}$	&  1  &  0  & 0 & E \cr
$A_m^{as}$	& 1/2 & 1/2 & 1 & C \cr 
$A_m^{st}$	&  0  &  0  & 1 & B \cr \hline
$C^{abc}$	& 1/2 & 1/2 & 0 & D \cr 
$C^{sab+}$	&  0  &  1  & 1 & B \cr
$C^{sab-}$	&  1  &  0  & 1 & A \cr
$C^{ast}$	& 1/2 & 1/2 & 1 & C \cr
$C^{stu}$	&  0  &  0  & 0 & F \cr \hline
$D^a$		& 1/2 & 1/2 & 0 & D \cr 
$D^s$		&  0  &  0  & 1 & A \cr \hline\hline
$\bar{F}^{ab}_{mn}$& 0&  1  & 0 &   \cr \hline
\end{tabular}
}
{Internal quantum numbers of the bosonic fields.}

The linearized equations of motion in $D=4$ are given by

\group{A}
\be
\del^2 B_{(st)} = \del^2 C^{sab-} = \del^2 D^s = \del^2(3B^a_a + 2 B^s_s) = 0.
\ee

\group{B}
\bea
\del^2 C^{sab+} &=& \half \bar{F}^{ab}_{mn}W^s_{mn} 
-\recip{ 4} \ep^{klmn}\bar{F}^{ab}_{kl} F^s_{mn}, \\
\del^n F^s_{nm} &=& -\recip{ 4}\ep{_m}^{nkl} \bar{F}^{ab}_{nk} \del_l C^{sab+} 
\  \ (F^s_{mn} \equiv \half \ep^{stu} F^{tu}_{mn} ), \\
\del^n W^s_{nm} &=& \half \bar{F}^{ab}_{mn}\del^n C^{sab+}. 
\slabel{grpD}
\eea

\group{C}
\bea
\del^2C^{as} &=& \recip{ 4} \ep^{klmn}\bar{F}^{ab}_{kl} F^{bs}_{mn} \  \ 
(C^{as} \equiv \half\ep^{stu}C^{atu}), \\
\del^2 B^{as} &=& \half\bar{F}^{ab}_{mn} F^{bs}_{mn},  \\
\del^nF^{as}_{nm} &=& - \bar{F}^{ab}_{mn}\del^n B^{bs} 
-\half \ep{_m}^{nkl} \bar{F}^{ab}_{nk} \del_l C^{bs}. 
\eea

\group{D}
\bea
\del^2 C^{abc} &=& {3\over 2} W^{[a}_{mn}\bar{F}^{bc]}_{mn}, \\
\del^2 D^a &=& \recip{ 4} \ep^{klmn}\bar{F}^{ab}_{kl}W^b_{mn}, \\
\del^nW_{nm}^a &=& -\half \ep{_m}^{nkl}\bar{F}^{ab}_{nk}\del_lD^b 
+ \half \bar{F}^{bc}_{mn}\del^nC^{abc}.  
\eea

\group{E}
\bea
\del^2 B^{(ab)} &=& 
\half F^{ac-}_{mn}\bar{F}^{cb}_{mn} + \half F^{bc-}_{mn}\bar{F}^{ca}_{mn}
+\half \bar{F}^{ac}_{mn}\bar{F}^{bd}_{mn} B^{(cd)}, \\
\del^n F^{ab-}_{nm} &=& 
\bar{F}^{bc}_{mn}\del^n B^{(ca)} - \bar{F}^{ac}_{mn} \del^n B^{(cb)}. 
\eea

\group{F}
\bea
\del^2 C &=& \recip{ 8} \ep^{klmn}\bar{F}^{ab}_{kl} F^{ab+}_{mn}, \\
\del^2 B^s_s &=& \recip{ 2} \bar{F}^{ab}_{mn} F^{ab+}_{mn} 
-\half \bar{F}^{ab}_{mk} \bar{F}^{ab}_{ml} h^{kl}, \\
\del^n F^{ab+}_{nm} &=& 
\bar{F}^{ab}_{kl} \del_k h_{lm} 
- \half \bar{F}^{ab}_{mn} (2\del^k h_{kn} - \del_n h^k_k - \del_n B^s_s)
+\half\ep{_m}^{kln}\bar{F}^{ab}_{kl} \del_n C, \\
R^{(1)}_{mn}(h) 
&=& \half\bar{F}^{ab}_{mk} F^{ab+}_{nk} + \half\bar{F}^{ab}_{nk} F^{ab+}_{mk}
-\recip{ 6} g_{mn} \bar{F}^{ab}_{kl} F^{ab+}_{kl} \xx
&& -\half \bar{F}^{ab}_{mk}\bar{F}^{ab}_{nl} h^{kl} 
+\recip{ 6} g_{mn} \bar{F}^{ab}_{kj} \bar{F}^{ab}_{kl} h^{jl}
+\recip{ 4}\bar{F}^{ab}_{mk}\bar{F}^{ab}_{nk} B^s_s -\recip{ 6} \del^2 B^s_s.
\eea

\subsection{Computation of mass spectrum in each multiplet}

We have separated fields which decouple from one another using their 
internal quantum numbers. We should now disentangle the field equations 
further and find out which field belong to which $\CN=2$ multiplet.
Obviously, the bosonic fields in the same multiplet satisfy the same 
field equations, and the same for the fermions.

In this section, we jump to solve the equations of motion of fields 
in each multiplet, except for the gravity multiplet which 
has been analyzed in detail in section \ref{toysec}.
The reduction of the equations obtained 
in the previous subsection to the final form require somewhat lengthy 
algebra, and we put it off until the next subsection.
The complication partly arises from the fact that we chose a specific 
$D=11$ configuration from the beginning. Although the M theoretic origin 
of the geometry is manifest in this framework, the U-duality invariance 
of the $D=4$ theory and its symmetry breaking pattern is obscured.
A manifestly duality invariant approach sketched in \cite{ads3bbb} 
could simplify the process to a large extent.

\subsubsection{Hyper multiplet}  

Minimally coupled scalars in $D=4$ belong to this multiplet.
Clearly, the KK modes have $m^2=j(j+1)$. It follows that $h=j+1$. 
There is no gauge symmetry associated with the scalars.

\subsubsection{Vector multiplet}

A vector multiplet contains a vector $A_m$ and two real scalars 
$\phi_1, \phi_2$. 
In the simplest case, $\phi_1$ couples to $A_\a$ only and 
$\phi_2$ to $A_\m$. 
The field equations for the first group are
\bea
(\del_x^2 +\del_y^2 - 2) \phi_1 &=& \half\ep^{\a\b}F_{\a\b}, \\
\del^mF_{m\a} &=& 4\ep_{\a\b}\del^\b \phi_1.
\eea
In the same gauge as in the toy model, $A_\a$ is expanded in 
the spherical harmonics as 
\be
A_\a = b^I \ep_{\a\b} \del^\b Y^I.
\ee
We then get the equations 
\be
\pmatrix{ \del_x^2 - j(j+1) -2 & - j(j+1) \cr -4 & \del_x^2 - j(j+1) }
\pmatrix{\phi_1 \cr b} = 0.  
\ee
along with the constraint 
\be
\del_\m A^\m =0
\ee
For $j\ge 1$, one finds that the mass eigenvalues are 
\be
m^2 = j(j-1), (j+1)(j+2)\  \ \imp \  \ h = j, j+2.
\ee
For $j=0$, $b$ is absent and $\phi_1$ has $m^2 = 2, h=1$.

The field equations for $\phi_2$ and $A_\m$ are
\bea
(\del_x^2 + \del_y^2 + 2)\phi_2 &=& \half\ep^{\m\n} F_{\m\n}, \\ 
\del^n F_{n\m} &=& 4\ep_{\m\n} \del^\n\phi_2. 
\eea
As in the toy model, one can use the constraint $\del^\m A_\m =0$ to set
$A_\m = \ep_{\m\n}\del^\n a$. The equations then become,
\be
\pmatrix{ \del_x^2 - j(j+1) +2  & - \del_x^2 \cr -4 & \del_x^2 - j(j+1) }
\pmatrix{\phi_2 \cr a} = 0.  
\ee
For $j\ge 1$, one finds the same mass eigenvalues as for $\phi_1$ and $A_\a$.
For $j=0$, $A_\m$ is a gauge field in $D=2$ and can be eliminated,
leaving $\phi_2$ with $m^2 = 2, h=1$.

\subsubsection{Gravitino multiplet}

Minimally coupled vectors in $D=4$ belong to this multiplet.
One obtain two $D=2$ scalars with $m^2 =j(j+1)$ for all $j\ge 1$. 
For $j=0$, the mode for $A_\a$ is absent and $A_\m$ becomes a gauge field 
in $D=2$, so there is no bulk degree of freedom.

\subsubsection{Gravity multiplet}

This multiplet was analyzed for the toy model case. Conformal weights of the bosonic states satisfy $h=j, j+2$. 

\subsection{Grouping $D=4$ fields into $\CN=2$ multiplets}

\group{A}

All the fields in this group are minimally coupled scalars in $D=4$ and 
belong to the hypermultiplet.

\group{B}

It is convenient to dualize $F^s_{mn}$ by defining
\be
\tilde{F}^s_{mn} \equiv 
-\half \ep{_{mn}}^{kl}F^s_{kl} - \bar{F}^{ab}_{mn}C^{sab+}. 
\ee
Then the equation of motion of $F^s$ becomes the Bianchi identity for 
$\tilde{F}^s$, and the Bianchi identity for $F^s$ become
\be
\del^n \tilde{F}^s_{nm} = \bar{F}^{ab}_{mn} \del^n C^{sab+}.
\ee
Note that the right-hand side is the same as that of \eq{grpD}. 
In terms of $\tilde{F}^s$, the field equation for $C$ becomes
\be
\del^2 C^{sab+} = \half\bar{F}^{ab}_{mn}
(\tilde{F}^s_{mn}+W^s_{mn} + \bar{F}^{cd}_{mn}C^{scd+}).
\ee
Clearly, $\tilde{F}^s - W^s$ decouple from $C$ and contribute to the 
$3/2$ multiplet. Since $C$ is coupled to $\tilde{F}^s + W^s$ 
by $\bar{F}^{ab}$, we find that $C^{s47+}$ decouples and contribute to
the hypermultiplet. 
The remaining fields belong to the vector multiplet. 
In particular, $C^{s45+}$ couples to $\tilde{A}^s_\m + V^s_\m$ and 
$C^{s46+}$ couples to $\tilde{A}^s_\a + V^s_\a$

\group{C}

Writing down all the components of the field equations and collecting those
which couple to one another, one finds twelve identical copies of the
following set of coupled equations:
\bea
\del^2 B &=& -\del^2 C = \half(-\ep^{\m\n} F_{\m\n} + \ep^{\a\b} F_{\a\b}),\\
\del^n F_{n\m} &=&  \ep_{\m\n} \del^\n (-B+C), \\
\del^n F_{n\a} &=& \ep_{\a\b} \del^\b (B-C). 
\eea
As before, we set,
\be
A_\a = b^I\ep_{\a\b}\del^\b Y^I,\  \ 
A_\m = A_\m^IY^I,\  \ A_\m = \ep_{\m\n} \del^\n a.
\ee
It is then easy to show that $(B+C)$ belong to the hypermultiplet, 
$(a+b)$ to the gravitino multiplet, and $(B-C)$ and $(a-b)$ together to 
the vector multiplet.

\group{D}

If we define
$$
C^a = \recip{3!} \ep^{abcd} C^{bcd},
$$
we find that the field equations have exactly the same structure as those in 
the previous group.

\group{E}

First, $B_{45}+B_{67}$, $B_{46} - B_{75}$ and $B_{44}-B_{55}-B_{66}+B_{77}$
decouple and contribute to the hyper multiplet. 
The other equations fall into six groups each of which contains one scalar and
one vector. Each group gives the spectrum for half a vector multiplet. 
Explicitly, the six groups are
\be
\begin{array}{cc}
(B^{44}+B^{55}-B^{66}-B^{77}, A_\m^{45-}), & 
(B^{44}-B^{55}+B^{66}-B^{77}, A_\a^{46-}), \cr
(B^{46}+B^{75}, A_\m^{47-}), & (B^{45}-B^{67}, A_\a^{47-}), \cr
(B^{47}-B^{56}, A_\m^{46-}), & (B^{47}+B^{56}, A_\a^{45-}). \cr
\end{array}
\ee

\group{F}

First, $h_2 \equiv {h^\a}_\a$, $B^s_s$, $A_\m^{45+}$ and $A_\a^{46+}$
belong to the gravity multiplet.
Second,  $h_{\m\a}$, $A_\m^{46+}$ and $A_\a^{45+},C$ 
belong to the vector multiplet. 
Finally, $A_\m^{47+}$ and $A_\a^{47+}$ decouple and 
contribute to the gravitino multiplet.

\mytable{bosonspec}{
\begin{tabular}{c|c|c|c|c|c} \hline
Group & grav & 3/2 & vec & hyper & Total \cr \hline
A & 0 &  0 &  0 & 18 & 18 \cr 
B & 0 &  4 &  8 &  4 & 16 \cr
C & 0 & 12 & 24 & 12 & 48 \cr
D & 0 &  6 & 12 &  3 & 21 \cr
E & 0 &  0 & 12 &  3 & 15 \cr
F & 4 &  2 &  4 &  0 & 10 \cr \hline
Sum &4& 24 & 60 & 40 & 128 \cr \hline
\end{tabular}
}
{\small This table summarizes the number of degrees of freedom a group of 
bosonic equations contribute to each of the four multiplets.}


\section{Fermionic mass spectrum}

\subsection{Linearized field equations and reduction to ${\mathbf D=4}$}

The linearized field equation for the gravitino in $D=11$ reads
\be
\G^{IJK}\del_J\Psi_K = 
\recip{ 4\cdot 4!} \G^{IJKLMN} \Psi_J F_{KLMN}
+\recip{ 8} \G^{JK}\Psi^L F{^I}_{JKL}. 
\ee
Throughout this section we suppress the bar on the background field strength.
The linearized local SUSY transformation law plays the role of gauge symmetry 
for Fermions:
\be
\d \Psi_M = \del_M\ep +
\recip{ 12\cdot 4!}F_{IJKL}(8\d^I_M\G^{JKL}-\G{^{IJKL}}_M)\ep.
\ee
In dimensional reduction to $D=4$, it is convenient to define
\be
\begin{array}{rcl}
\l = \G^a \Psi_a,&& \Psi_{(a)} = \Psi_a - \recip{ 4} \G_a \l, \\
\chi = \G^s \Psi_s,&& \Psi_{(s)} = \Psi_s - \recip{ 3} \G_a \chi.
\end{array}
\ee
The following shift in the $D=4$ spin $3/2$ fields bring their kinetic term 
into the standard form.
\be
\Psi_m^{(11)} = \Psi_m^{(4)} -\half\G_m(\l+\chi).
\ee
We then decompose the fermion into chiral and anti-chiral components with respect to $SO(4)$ of $T^4$,
\be
\Psi^{\pm} \equiv {1 \over 2}(1 \pm \bar \Gamma) \Psi
\ee
where $\bar \Gamma \equiv {1 \over 4!} \ep_{abcd} \Gamma^{abcd}$. 
As in the previous section, we can divide the field equations into a few groups
using the internal symmetry. 
After some gamma matrix algebra, one finds that the field equations 
and gauge transformation laws in $D=4$ are given by

\mytable{fermiongrp}
{
\begin{tabular}{c|cc|c|c}
\hline
Field	        &$SU(2)_+$&$SU(2)_-$&$SU(2)_3$& Group \cr \hline
$\Psi_m^+$	& 1/2 &  0  & 1/2 & I \cr
$\Psi_m^-$	&  0  & 1/2 &     & J \cr \hline
$\Psi_{(a)}^+$	&  1  & 1/2 & 1/2 & H \cr
$\Psi_{(a)}^-$ 	& 1/2 &  1  &     & I \cr \hline
$(3\l+2\chi)^+$	& 1/2 &  0  & 1/2 & G \cr 
$(3\l+2\chi)^-$	&  0  & 1/2 &     & H \cr \hline
$\Psi_{(s)}^+$	& 1/2 &  0  & 3/2 & G \cr 
$\Psi_{(s)}^-$	&  0  & 1/2 &     & H \cr \hline 
$\chi^+$	& 1/2 &  0  & 1/2 & I \cr
$\chi^-$	&  0  & 1/2 &     & J \cr \hline 
\end{tabular}
}
{Internal quantum numbers of the fermionic fields.}

\group{G}
\bea
\G^n\del_n \Psi_{(s)}^+ = \G^n\del_n(3\l+2\chi)^+ &=& 0,\\
\d \Psi_{(s)}^+ = \d (3\l+2\chi)^+ &=& 0.
\eea

\group{H}
\bea
\G^n\del_n \Psi_{(a)}^+ &=& \recip{ 4} F^{ad}_{ij}\G^{ij}\Psi^+_{(d)}, \\
\G^n\del_n(3\l+2\chi)^- &=& 
\recip{ 16}F^{cd}_{ij}\G^{ij}\G^{cd}(3\l +2\chi)^-,\\
\G^n\del_n \Psi_{(s)}^- &=& 
-\recip{ 16}F^{cd}_{ij}\G^{ij}\G^{cd}\Psi_{(s)}^-, \\
\d\Psi_{(a)}^+ &=& \d (3\l+2\chi)^- = \d \Psi_{(s)}^- = 0.
\eea

\group{I}
\bea
\G^n\del_n\Psi_{(a)}^- &=& 
-\recip{ 8}F^{ab}_{ij}\G^b
\{\G^{ij} \chi^+ + (\G^{ijk}-2\G^i\d^{jk})\Psi_k^+ \}, \\
\G^{mnk}\del_n\Psi_k^+ &=&
-\recip{ 8}F^{ab}_{ij}(\G^{mij}-2\d^{mi}\G^j) \G^a\Psi_{(b)}^-, \\
\G^n\del_n \chi^+ &=& \recip{ 4} F^{ab}_{ij} \G^{ij}\G^a\Psi_{(b)}^-, \\
\d \Psi_{(a)}^- &=& \recip{ 8} F^{ab}_{ij}\G^b\G^{ij} \ep^+,\\
\d \Psi_m^+ &=& \del_m \ep^+,\  \ \d \chi^+ = 0.
\eea

\group{J}
\bea
\G^n\del_n\chi^- &=& 
-\recip{ 16}F^{cd}_{ij}\G^{cd}(\G^{ijk}-2\G^i\d^{jk})\Psi^-_k, \\
\G^{mnk}\del_n\Psi_k^- &=& 
\recip{ 16} (F_{ij}^{ab} \G^{ijmn} +2F_{mn}^{ab}) \G^{ab}\Psi_n^- 
+\recip{ 32} F^{ab}_{ij}(\G^{mij}-2\d^{mi}\G^j)\G^{ab}\chi^-, \\
\d \Psi_m^- &=& 
\del_m\ep^- - \recip{ 32}F^{ab}_{ij}(\G^{mij}-2\d^{mi}\G^j)\G^{ab}\ep^-, \\
\d\chi^- &=& -\recip{ 16} F^{ab}_{ij}\G^{ab}\G^{ij}\ep^-.
\eea

\subsection{Computation of the mass spectrum in each multiplet} 

\subsubsection{Hyper multiplet}

Spinors with nonzero mass generated by the background gauge field 
belong to this multiplet. After diagonalizing the mass matrix, they 
satisfy the equation of motion of the form
\be
\G^m\del_m \psi + i\G^{01} \psi = 0.
\ee
One finds that $|m| = j+\half \pm 1$, which implies that $h=j, j+2$. 
There is no gauge symmetry acting on the spinors in this multiplet.

\subsubsection{Vector multiplet}

Minimally coupled massless spinors in $D=4$ belong to this multiplet. 
One easily finds that $h = |m| + \half = j+1$ for all $j\ge \half$. 
There is no gauge symmetry acting on the spinors in this multiplet. 

\subsubsection{Gravitino multiplet}

A gravitino multiplet contains a gravitino $\psi$ and a spinor $\chi$. 
In the same notation as in the toy model, their coupled equations of motion
break up as follows
\bea
(\dirac_x + \bar{\ga}\dirac_y) \chi &=& -i\bar{\ga} (\eta -\l),\slabel{1st} \\
(\dirac_x + \bar{\ga}\dirac_y) \eta +  \bar{\ga}\dirac_y \l
-2 \del^\a \psi_{(\a)} &=& -i\bar{\ga} \chi, \slabel{2nd} \\
(\dirac_x + \bar{\ga}\dirac_y) \lambda + \dirac_x \eta
-2 \del^\m \psi_{(\m)} &=& i\bar{\ga} \chi, \slabel{3rd} \\
-\del_{(\m)}\eta + \bar{\ga} \dirac_y \psi_{(\m)} &=& 0, \slabel{4th} \\
-\del_{(\a)}\l + \dirac_x \psi_{(\a)} &=& 0.\slabel{5th}
\eea
where we expressed the four dimensional gamma matrices as tensor products of two dimensional ones as in the case of the toy model, and  $\dirac_x, \dirac_y$ are the two dimensional dirac operators. 
The gauge transformation laws are given by
\be
\begin{array}{lll}
\d \psi_{(\m)} = \del_{(\m)} \ep, & \d \l = \dirac_x \ep, & \cr
\d \psi_{(\a)} = \del_{(\a)} \ep, & \d \eta = \bar{\ga} \dirac_y \ep,& 
\d \chi = -i\bar{\ga} \ep.
\end{array}
\ee
One can always gauge away $\eta$. Then \eq{4th} sets $\psi_{(\m)}$ to zero.

For $j\ge 3/2$, \eq{2nd} determine $\psi_{(\a)}$ algebraically. The only
independent equations that remain are \eq{1st} and \eq{3rd} with 
$\eta$ and $\psi_{(\m)}$ removed. The mass eigenvalues are the same as 
those of hyper multiplet: $|m| = j+\half \pm 1$, $h = j, j+2$. 
For $j=1/2$, the modes for $\psi_{(\a)}$ are absent and \eq{5th} is trivially
satisfied. Eq. \eq{2nd} gives an algebraic relation between $\l$ and $\chi$.
So the number of degrees of freedom is reduced by half. 
One finds $h=j+2$ for all modes.

\subsubsection{Gravity multiplet}
The equations satisfied by this multiplet was analyzed for the toy model case. One finds $h=j+1$, with number of degrees reduced by half for $j=1/2$.

\subsection{Grouping $D=4$ fields into $\CN=2$ multiplets}

\group{G}

All the spinors in this group are massless and minimally coupled in $D=4$ and belong 
to the vector multiplet. 

\group{H}

We choose the following basis for $SO(4)$ gamma matrices,
\be
\G^4 = \pmatrix{0&\identity \cr \identity&0},\  \ 
\G^{5,6,7} = i\pmatrix{0&-\s^{1,2,3} \cr \s^{1,2,3}&0},
\ee
where $\s^i$ are the Pauli matrices.
In this basis, an $SO(4)$ spinor splits into two chiral spinors as
\be
\eta = \pmatrix{\eta^- \cr \eta^+}.
\ee
Consider $(3\l+2\chi)^-$ first. To simplify the equations, we use the letter
$\Psi$ to denote $(3\l+2\chi)^-$ in the equations to follow.
In the basis we chose, the field equation reduces to
\be
\G^n\del_n \Psi = \textstyle{i\over 2} 
(\s^1 \otimes \G^{01} + \s^2 \otimes \G^{23})\Psi. 
\ee
We can further decompose the equation by setting 
\be
\Psi = \pmatrix{\Psi_1 \cr \Psi_2}.
\ee
After splitting each of $\Psi_{1,2}$ into two pieces according to their 
chirality in the non-compact $D=4$ spacetime, and recombining those pieces
which couple to each other, one finds that one linear combination 
belongs to the vector multiplet and the other one to the hyper multiplet.

The spectrum is exactly the same for $\Psi_{(s)}^+$ except that it has
twice as many degrees of freedom as $3\l+2\chi$. We find the same result 
even for $\Psi_{(a)}^+$ again except for the degeneracy. 
In counting the degeneracy, one should remember the constraint 
$\G^a\Psi_{(a)} = 0$. In the basis we chose above, it reduces to
\be
\Psi_{(4)} - i\s^1 \Psi_{(5)} -i\s^2 \Psi_{(6)} -i\s^3 \Psi_{(7)} = 0.
\ee

\group{I}

Doing the same sort of recombination of spinors as above, one finds the 
following results.
\begin{enumerate}

\item 
A third of $\Psi_{(a)}^-$ decouple from all the other fields. 
They belong to the vector multiplet.

\item
Another third of $\Psi_{(a)}^-$ couple to $\chi^+$. 
They contribute to the hypermultiplet.

\item 
The last third of $\Psi_{(a)}^-$ couple to $\Psi_m^+$. 
They belong to the spin $3/2$ multiplet.

\end{enumerate}

\group{J}

\begin{enumerate}

\item
A half of $\chi^-$ decouple and belong to the vector multiplet.

\item
A half of $\Psi_m^-$ decouple and satisfy the same equation as the gravitino 
in the toy model. Therefore they belong to the gravity multiplet.

\item
The other components of $\chi^-$ and $\Psi_m^-$ couple to each other.
They contribute to the spin $3/2$ multiplet.

\end{enumerate}

The following table summarizes the result of this section.

\mytable{fermispec}
{
\begin{tabular}{c|c|c|c|c|c} \hline
Group & grav & 3/2 & vec & hyper & Total \cr \hline 
G & 0 &  0 & 24 &  0 & 24 \cr 
H & 0 &  0 & 24 & 24 & 48 \cr
I & 0 & 16 &  8 & 16 & 40 \cr
J & 4 &  8 &  4 &  0 & 16 \cr \hline
Sum& 4& 24 & 60 & 40 & 128\cr \hline
\end{tabular}
}
{Summary of the fermionic spectrum}




\vskip 1cm
\centerline{\bf ACKNOWLEDGEMENT}
\vskip 0.5cm
We are grateful to Seungjoon Hyun for many helpful discussions, 
Youngjai Kiem for carefully reading the manuscript and 
making useful comments, and Jan de Boer for a correspondence.
SL would like to thank Jeremy Michelson for discussions at Carg\`ese '99 ASI.

\appendix

\section{Notations and Conventions \label{conv}}

We consider $D=11$ SUGRA on $AdS_2\times S^2\times T^4\times T^3$. 
Each manifold in the product is parametrized by 
$x^\m$ $(\m=0,1)$, $y^\a$ $(\a=2,3)$, $z^a$ $(a=4,5,6,7)$ 
and $w^s$ $(s=8,9,10)$, respectively. 
We use the indices $(M,N,\cdots)$ to label all eleven coordinates together 
and $(m,n,\cdots)$ to label the coordinates of $AdS_2\times S^2$.
The signature of the metric is $(-+ \cdots +)$.
The field strength of a $p$-form potential in any dimension is defined by
\be
F_{M_0 \cdots M_p} = p \del_{[M_0}A_{M_1 \cdots M_p]}.
\ee

\section{Spherical Harmonics \label{sphh}}

The spherical harmonics form a basis for the fields living on a sphere.  
In this appendix we consider only the case of $S^2$. We can construct them 
by considering the eigenstates of maximal commuting subalgebra of 
$SU(2)$ group, which are the total angular momentum $\vec J^2=j(j+1)$, 
its $z$ component $J_z=m$, the orbital angular momentum $\vec L^2= l(l+1)$, 
and the spin $\vec S^2 = s(s+1)$. 
The case for the scalar is easiest since $s=0$ and $\vec J^2 = \vec L^2$, 
which we identify with the Laplacian on the sphere, $\nabla_y^2$, 
where $y$ indicates two dimensional coordinates parametrizing the two-sphere. 
This expression for $\vec L^2$ can  be obtained by embedding $S^2$ into 
three dimensional space, writing down $\vec L^2$ in terms of the 
Cartesian coordinates, which is quite well known, and 
reexpressing these in terms of polar coordinates. 
Therefore, by construction, we have
\be
\nabla_y^2 Y^{(j,m)}(y) = -j(j+1) Y^{(j,m)}(y) \label{reo},
\ee
where $Y^{(j,m)}(y)$ denotes the eigenstates with eigenvalues $(j,m)$, 
which were defined above.

Next consider the spinor spherical harmonics where $s=1/2$. 
The easiest way to consider it is to embed the sphere in the 
three dimensional space and use cartesian coordinates. 
We construct them by taking the tensor product of scalar spherical harmonics 
with 2-component spinor, and taking appropriate linear combinations. 
We then get the expression\cite{sakurai} 
\be
\Sigma^{j=l\pm 1/2,m}_l
= {1 \over \sqrt{2l +1}} \left( 
\matrix{\pm \sqrt{\l \pm m + {1 \over 2} }Y^{m-1/2}_l (\theta, \phi) \cr 
\sqrt{\l \mp m+ {1 \over 2} }Y^{m-1/2}_l (\theta, \phi) } \right),
\ee
where the labels indicates the eigenvalues as usual, and all the harmonics 
are normalized to unity. For given $j$, the only possible values of $l$ are 
$j \pm {1 \over 2}$, so the degeneracy is $2(2j +1)$.
However, it is convenient for our purpose to group the spherical harmonics of 
given $j$ according to the eigenvalue of $\tilde J \equiv \dirac _y$ 
rather than $m, l$, where $\dirac_y \equiv \tau^\a \nabla_\a$ 
is the two dimensional dirac operator on the sphere with $\tau_\a$ 
given by the usual Pauli matrices.  One can show that 
\be
\vec J^2 = \dirac_y^2 - {1 \over 4}, 
\ee
by comparing the two dimensional operator with the expression 
in the embedding three dimensional cartesian coordinates, 
so it is obvious that $\tilde J$ commutes with $\vec J^2$, 
and its eigenvalues are $\pm (j+{1 \over 2})$.  
However, it turns out that neither of $\vec L^2$ nor $J_z$ commutes with 
$\tilde J$.  Therefore we have to find two other operators which commute with 
$\vec J^2, \tilde J$ in order to distinguish linearly independent 
spherical harmonics. We will not identify them since they are not needed 
for the present discussion. Note that the chirality operator 
$\bar{\tau} \equiv {1 \over 2}\ep_{\a \b} \tau^\a \tau^\b$ 
anticommutes with $\tilde J$. Therefore  given an eigenstate with  
$\tilde J>0$, which we denote by $\Sigma^j_+$, we have a counterpart 
$\Sigma^j_- \equiv \bar \tau \Sigma^j_+$ and vice versa. 
\footnote{ Our notations closely mimic those in \cite{ads5a}, 
but the convention for the $\pm$ sign is flipped.} 
This immediately implies that there are same number of $\Sigma^j_+$ states 
and $\Sigma^j_-$ states for given $j$. 
Since the degeneracy of total states is $2(2j +1)$, we have $2j+1$ 
$\Sigma^j_+$ (or $\Sigma^j_-$ ) states.

We also state without proof that the lowest spinor spherical harmonics are 
killing spinors, satisfying the relation
\be
(\nabla_\alpha \mp {i \over 2}\tau_\a) \Sigma^{1/2}_\pm = 0.  \label{reth}
\ee


\newpage


\small\normalsize

\begin{thebibliography}{99}

\bibitem{malda}
J. Maldacena, 
``The Large $N$ limit of Superconformal Field Theories and Supergravity,''
\atmp{2}{1998}{231}, hep-th/9711200.

\bibitem{gkp}
S. S. Gubser, I. R. Klebanov and A. M. Polyakov,
``Gauge Theory Correlators from Non-Critical String Theory,''
\pl{428}{1998}{105}, hep-th/9802109.

\bibitem{witt}
E. Witten,
``Anti de Sitter Space and Holography,''
\atmp{2}{1998}{253}, hep-th/9802150.

\bibitem{rev}
For a review, see \\
O. Aharony, S. S. Gubser, J. Maldacena, H. Ooguri and Y. Oz,
``Large $N$ Field Theories, String Theory and Gravity,''
hep-th/9905111.

\bibitem{gibtown}
G. Gibbons and P. K. Townsend, 
``Black Holes and Calogero Models,'' 
hep-th/9812034.

\bibitem{town}
P. K. Townsend,
``The M(atrix) Model/$AdS_2$ Correspondence,''
Procedings of the 3rd Puri workshop on Quantum Field Theory,
hep-th/9903043.

\bibitem{strom}
A. Strominger, 
``$AdS_2$ Quantum Gravity and String Theory,'' 
hep-th/9809027.

\bibitem{frag}
J. Maldacena, J. Michelson and A. Strominger,
``Anti-de Sitter Fragmentation,''
hep-th/9812073.

\bibitem{nak}
T. Nakatsu and N. Yokoi, 
``Comments on Hamiltonian Formalism of AdS/CFT Correspondence,''
hep-th/9812047.

\bibitem{cad}
M. Cadoni and S. Mignemi,
``Asymptotic Symmetries of $AdS_2$ and Conformal Group in $d=1$,''
hep-th/9902040.

\bibitem{tsey}
I. R. Klebanov and A. A. Tseytlin,
``Intersecting M-branes as Four-dimensional Black Holes,''
\np{475}{1996}{179}.

\bibitem{berob}
B. Bertotti, \pr{116}{1959}{1331}; \\
I. Robinson, Bull. Acad. Polon. Sci. {\bf 7} (1959) 351.

\bibitem{ads2a}
A. Fujii and R. Kemmoku,
``$D=5$ Simple Supergravity on $AdS_2\times S^3$,''
hep-th/9903231.

\bibitem{ads3a}
S. Deger, A. Kaya, E. Sezgin and P. Sundell,
``Spectrum of $D=6$, $N=4b$ Supergravity on AdS$_3\times S^3$,''
hep-th/9804166.

\bibitem{ads3aa}
A. Fujii, R. Kemmoku and S. Mizoguchi,
``$D=5$ Simple Supergravity on $AdS_3\times S^2$ and $\CN=4$ 
Superconformal Field Theory,''
hep-th/9811147.

\bibitem{ads4a}
 Castellani, R. D'Auria, P. Fr\'e, K. Pilch and P. van Nieuwenhuizen,
``The Bosonic Mass Formula for Freund-Rubin Solutions of $d=11$ 
Supergravity on General Coset Manifolds,''
\cqg{1}{1984}{339}.

\bibitem{ads5a}
H. J. Kim, L. J. Romans and P. van Nieuwenhuizen,
``Mass Spectrum of Chiral Ten-dimensional $N=2$ Supergravity on $S^5$,''
\prd{32}{1985}{389}.

\bibitem{ads7a}
P. van Niewenhuizen,
``The Complete Mass Spectrum of $d=11$ Supergravity Compactified on $S_4$
and a General Mass Formula for Arbitrary Cosets $M_4$,''
\cqg{2}{1985}{1}.

\bibitem{ads3b}
J. de Boer, 
``Six-dimensional Supergravity on $S^3\times$AdS$_3$ 
and 2d Conformal Field Theory,''
hep-th/9806104.

\bibitem{ads3bb}
M. G\"unaydin, G. Sierra and P. K. Townsend,
``The Unitary Supermultiplets of $d=3$ Anti-de Sitter and 
$d=2$ Conformal Superalgebras,''
\np{274}{1986}{429}.

\bibitem{ads3bbb}
F. Larsen,
``The Perturbation Spectrum of Black Holes in $\CN=8$ Supergravity,''
\np{536}{1998}{258}, hep-th/9805208.

\bibitem{ads4b}
M. G\"unaydin and N. P. Warner,
``Unitary Supermultiplets of $Osp(8/4,R)$ and 
the Spectrum of the $S^7$ Compactification of $11$-dimensional Supergravity,''
\np{272}{1986}{99}.

\bibitem{ads5b}
M. G\"unaydin and N. Marcus,
``The Spectrum of the $S^5$ Compactification of the Chiral 
$N=2$, $D=10$ Supergravity and the Unitary Supermultiplets of $U(2,2/4)$,''
\cqg{2}{1985}{L11}.

\bibitem{ads7b}
M. G\"unaydin, P. van Niewenhuizen and N. P. Warner,
``General Construction of the Unitary Representations of Anti-de Sitter
Superalgebras and the Spectrum of the $S^4$ Compactification of 
$11$-dimensional Supergravity,''
\np{255}{1985}{63}.

\bibitem{osc1}
M. G\"unaydin and C. Sa\c{c}lio\v{g}lu,
``Bosonic Construction of the Lie Algebra of Some Noncompact Groups 
Appearing in the Supergravity Theories and their Oscillator-like 
Unitary Representations,''
\pl{108}{1982}{180}.

\bibitem{osc2}
M. G\"unaydin and C. Sa\c{c}lio\v{g}lu,
``Oscillator-like Unitary REpresentations of Non-compact Groups with a Jordan
Structure and the Non-compact Groups of Supergravity,''
\cmp{87}{1982}{159}.

\bibitem{osc3}
I. Bars and M. G\"undaydin,
``Unitary Representations of Non-Compact Supergroups,''
\cmp{91}{1983}{31}.



\bibitem{sakurai} 
J. J. Sakurai, 
``Modern Quantum Mechanics,'' 
Benjamin/Cummings Publishing Company, Inc.

\bibitem{jeremy}
J. Michelson and M. Spradlin, 
``Supergravity Spectrum on $AdS_2\times S^2$,''
hep-th/9906056.

\bibitem{corley}
S. Corley, ``Mass Spectrum of $\CN=8$ Supergravity on $AdS_2\times S^2$,''
hep-th/9906102.

\end{thebibliography}
\end{document}